\documentclass[letterpaper, 10 pt, conference]{ieeeconf}  
\IEEEoverridecommandlockouts                              
\usepackage{graphicx}
\usepackage{algorithm}
\usepackage{float}
\usepackage{booktabs}
\usepackage{array}
\usepackage{amsmath} 
\usepackage{url}
\usepackage{amsmath}
\makeatletter
\def\maketag@@@#1{\hbox{\m@th\normalfont\normalsize#1}}
\makeatother
\usepackage{algorithm}
\usepackage{setspace}
\usepackage{tikz}
\usepackage{lipsum}
\usepackage{comment}
\usepackage{fancyhdr}
\usepackage{mathtools}
\usepackage{algpseudocode}
\usepackage{setspace}

\fancypagestyle{firstpage}{\lhead{\vspace{-2.75 cm} \small \textbf{{\fontfamily{cmss}\selectfont
The 62nd Conference on Decision and Control (CDC), December 13--15, 2023, Singapore}}}}

\title{\Large \bf
Robust Model Predictive Control for Enhanced Fast Charging on Electric Vehicles through Integrated Power and Thermal Management 
}

\author{Qiuhao Hu$^{1}$, Mohammad Reza Amini$^{1}$, Ashley Wiese$^{2}$, Ilya Kolmanovsky$^{3}$ and Jing Sun$^{1}$
\thanks{$^{1}$Q. Hu, M.R. Amini, and J. Sun are with the Department of Naval Architecture and Marine Engineering, University of Michigan, Ann Arbor, MI 48109 USA. (e-mail: \tt\small \{qhhu;mamini;jingsun\}@umich.edu) }%
\thanks{$^{2}$A. Wiese is with Ford Motor Company, MI 48126, USA. (e-mail: \tt\small awiese@ford.com)}%
\thanks{$^{3}$I. Kolmanovsky is with the Department of Aerospace Engineering, University of Michigan, Ann Arbor, MI 48109, USA. (e-mail: \tt\small ilya@umich.edu)}%
}

\begin{document}

\maketitle
\thispagestyle{firstpage}
\pagestyle{empty}

\makeatletter
\AfterEndEnvironment{algorithm}{\let\@algcomment\relax}
\AtEndEnvironment{algorithm}{\kern2pt\hrule\relax\vskip3pt\@algcomment}
\let\@algcomment\relax
\newcommand\algcomment[1]{\def\@algcomment{\footnotesize#1}}
\renewcommand\fs@ruled{\def\@fs@cfont{\bfseries}\let\@fs@capt\floatc@ruled
  \def\@fs@pre{\hrule height.8pt depth0pt \kern2pt}%
  \def\@fs@post{}%
  \def\@fs@mid{\kern2pt\hrule\kern2pt}%
  \let\@fs@iftopcapt\iftrue}
\makeatother

\begin{abstract}
This paper explores the synergies between integrated power and thermal management (iPTM) and battery charging in an electric vehicle (EV). A multi-objective model predictive control (MPC) framework is developed to optimize the fast charging performance while enforcing the constraints in the power and thermal loops. The approach takes into account the coupling of the battery and cabin thermal management. The case study of a commercial EV demonstrates that the proposed method can effectively meet the requirements of fast charging and thermal management when accurate preview information is available. However, failure to predict the charging event can result in performance degradation with longer charging time. A time-varying weighting strategy is proposed to enhance charging performance in the presence of uncertainty. This strategy leverages the battery state-of-charge ($SOC$) and adjusts the priority of the multi-objective MPC at different phases during charging. Simulated results using a commercial EV use case show improved robustness in charging time using the proposed strategy.

\end{abstract}

\vspace{-4pt}
\section{INTRODUCTION}\label{section.1} 

Vehicle electrification is a critical strategy to address the pressing energy and climate change challenges. Electric vehicles (EVs) leverage clean and renewable energy sources, have the potential for lower environmental impact, and lower maintenance and operating costs than conventional vehicles~\cite{choi2018effect, kumar2020adoption}. While the past few years have witnessed a significant increase in EV popularity and market share, there remain several technical challenges that impeded widespread adoption. Particular relevant for this study, is the combination of limited driving range~\cite{lee2013rapid,chen2021charging}, relatively long charging times~\cite{rauh2015understanding}, and availability of charging infrastructure which motivates the optimization of EV charging performance. This can be achieved through advancements in battery technology, charging infrastructure, and thermal management systems. This paper aims to enhance the fast-charging performance of EVs through the optimization of integrated power and thermal management (iPTM).

The battery charging rate of electric vehicles at a given charging station is determined by various factors such as the battery's state-of-charge ($SOC$), temperature, charging station capability, and overall health~\cite{jaguemont2018thermal}. Focusing on thermal considerations, cold operating temperatures can reduce the maximum charging rate and extend the charging time~\cite{hamednia2022optimal,hamednia2022optimal2}. Similarly, charging rates may be constrained in order to keep battery temperatures within a desirable range ~\cite{zhou2019impedance}. Although many efforts have been extended to enhance charging performance~\cite{hoke2014accounting, zhang2014study}, there are only a few works~\cite{hamednia2022optimal,dahmane2021optimized} exploring the coupling of battery charging and its thermal behavior.

The reduction of charging capacity due to cold temperatures can have a negative impact on the performance of electric vehicle charging. To mitigate this impact on energy efficiency and charging time, an optimization-based battery thermal management strategy was developed based on nonlinear programming in~\cite{hamednia2022optimal}. The study demonstrated that properly preconditioning the battery temperature can lead to a significant reduction in charging time. Furthermore, the approach was extended in~\cite{hamednia2022optimal2} to consider the detour time to the charging station during long-distance trips. The simulation results indicated that there is a trade-off between trip time and energy consumption.

Unlike cold ambient conditions, where the restriction in the charging power and low battery temperature at the start of charge leads to longer charging time, hot ambient conditions can reduce the heat rejection capacity of the battery cooling system. If the heat generated during fast charging exceeds the cooling capacity, it can cause the charging rate to be throttled to maintain the battery temperature within acceptable limits, thereby slowing down the charging process. To address this issue, an MPC-based iPTM strategy was proposed in our previous work~\cite{QHu2023IFAC}. This work in~\cite{QHu2023IFAC}, however, does not consider cabin thermal management. In EV applications where the battery and cabin thermal management share one heat rejection system, their coupling needs to be properly considered, particularly when both battery and cabin require large heating/cooling demands. This study aims to extend the MPC-based iPTM by integrating the battery and cabin thermal management and exploring the synergy between iPTM and fast charging.  

The contributions of this paper are as follows: Firstly, we consider both battery and cabin thermal management and extend the MPC-based iPTM approach with a non-uniform sampling strategy and a shrinking horizon to optimize the fast charging performance. Secondly, the impact of uncertainty associated with the prediction of a charging event is investigated using a commercial EV use case, quantifying the trade-off between charging time and cabin comfort. Thirdly, a time-varying weighting strategy is proposed, which adjusts the penalty weights based on the current value of battery $SOC$ to enhance the charging performance in the presence of uncertainty.

\section{Models of Power and Thermal Systems of A Commercial Electric Vehicle} \label{section.2} 

While the approach presented in this paper can be applied to a broader class of EVs, we focus our case study on the electric vehicle for commercial use, referred to as commercial electric vehicle. In this section, a detailed description of the models used to represent the thermal and power subsystems of a commercial electric vehicle utilized in this study is provided.

\subsection{Battery State-of-Charging Model}\vspace{-0pt}
An equivalent circuit model~\cite{he2011evaluation} is used to represent the battery SOC dynamics.
\begin{gather}\label{eq:SOC_dynamics}
\dot{SOC}=f_{SOC}(t)=\frac{-I_{bat}}{C_{bat}}=-\frac{U_{oc}-{\sqrt {U_{oc}^2-4R_{int}P_{bat}}}}{{2{R_{int}{C_{bat}}}}},
\end{gather}
where $I_{bat}$ and $C_{bat}$ are the battery current and capacity, respectively. The battery current is determined by the battery power ($P_{bat}$), the open circuit voltage ($U_{oc}$), and the internal resistance ($R_{int}$). When the vehicle is in motion, the battery power is the sum of the vehicle traction power ($P_{trac}$) and the power consumed by auxiliary subsystems ($P_{aux}$), i.e., $P_{bat}=P_{trac}+P_{aux}$. While the vehicle is being charged at the station, the net charging power is $P_{bat}=P_{chg}+P_{aux}$. Note that $P_{bat}$ is negative.

\subsection{Battery Thermal Model}
The battery is modeled as a lumped thermal mass and its temperature dynamics are expressed using the following equation:
\vspace{-0.15cm}
\begin{gather}\label{eq:T_bat}
\dot{T}_{bat}=f_{bat}(t)=\frac{1}{m_{bat,thm}C_{bat}}(\dot{Q}_{gen}-\dot{Q}_{amb}-\dot{Q}_{bat}),
\end{gather}
where $m_{bat}$ and $C_{bat}$ are the battery thermal mass and specific heat capacity, respectively. $\dot{Q}_{gen}$ and $\dot{Q}_{amb}$ represent the irreversible heat rate generated by internal resistance and the heat exchange rate between the battery and ambient, respectively, which are expressed as:

\vspace{-0.15cm}
\begin{gather}\label{eq:T_bat}
\dot{Q}_{gen}=I_{bat}^2R_{int},\\
\dot{Q}_{amb}=\alpha_{amb}(T_{amb}-T_{bat}),
\end{gather}
where $\alpha_{amb}$ is a coefficient of heat exchange rate. Additionally, $\dot{Q}_{bat}$ represents the battery cooling power provided by the cooling system.

\vspace{-00pt}
\begin{figure}[th!]
	\begin{center}
	\includegraphics[width=0.95\columnwidth]{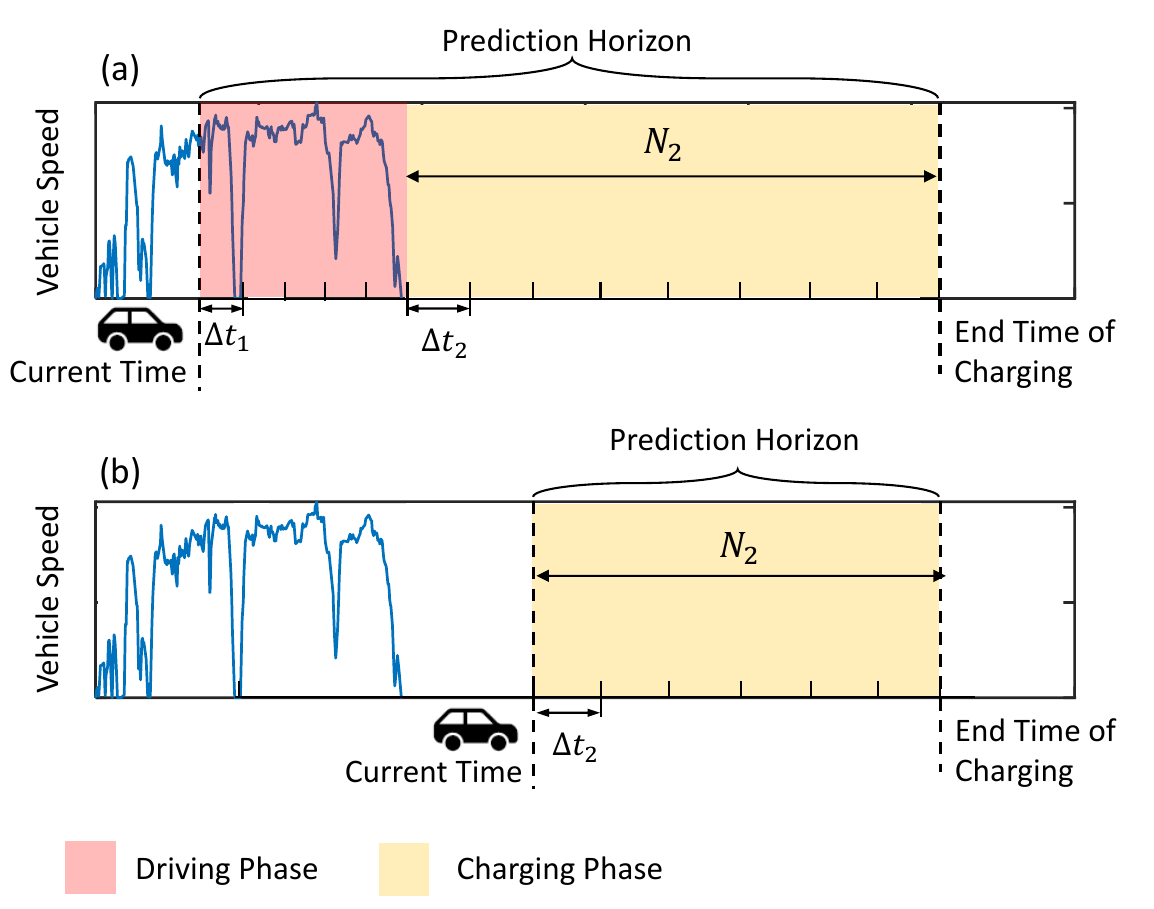}\vspace{-0pt}
		\caption{Concept of the developed MPC with nonuniform sampling and shrinking horizon: (a) driving phase, before the vehicle arrives at the charging station, and (b) charging phase, after the vehicle arrives at the charging station.}\vspace{-10pt}
	     \label{fig:Concept_MPC} 
	\end{center}
\end{figure}

\subsection{Cabin Thermal Model}
The cabin is also modeled as a lumped mass so that the cabin temperature dynamics are expressed as
\vspace{-0.10cm}
\begin{equation} \label{eq:T_cab}
\small
\begin{aligned}
&\dot{T}_{cab}=f_{cab}(t)\\
&=\frac{1}{m_{cab}C_{cab}}(\dot{Q}_{sun}+\dot{Q}_{cov}+\dot{Q}_{ven}+\dot{Q}_{met}-\dot{Q}_{cab}),
\end{aligned}
\end{equation}
where $m_{cab}$ and $C_{cab}$ are the thermal mass and specific heat capacity of cabin, respectively. $\dot{Q}_{sun}$, $\dot{Q}_{cov}$, $\dot{Q}_{ven}$, and $\dot{Q}_{met}$ are the heat transfer rate due to solar loading, air convection, air ventilation, and human metabolic activities, respectively. For a more detailed formulation of each heat source term in (\ref{eq:T_cab}), please refer to~\cite{fayazbakhsh2013comprehensive}. Moreover, $\dot{Q}_{cab}$ is the cabin cooling power provided by the cooling system.

\section{Model Predictive Control Formulation} \label{section.3}

Our analysis focuses on commercial EVs that operate at hot ambient temperatures with low initial $SOC$, which requires a fast-charging at a nearby charge station to complete the vehicle mission. The objectives of the iPTM are as follows:

\begin{enumerate}
\item final $SOC$ above a prescribed threshold;
\item total charging time within a desired range;
\item minimizing energy consumption while enforcing constraints on thermal states and control inputs.
\end{enumerate}

Note that for commercial vehicles, the final $SOC$ and total charging time requirements are typically determined by the missions after charging. The main challenge of applying a conventional receding-horizon MPC in such a setting is the variability in the $SOC$ and time required for charging, which makes it uncertain when exactly the target $SOC$ will be reached. Therefore, the time instant when the vehicle completes the charging can be within or beyond the receding prediction horizon. This makes the problem formulation different from the one addressed by the conventional MPC.

To address the aforementioned challenges and achieve the objectives, a novel MPC-based iPTM with a non-uniform sampling strategy and a shrinking horizon is developed. Fig.~\ref{fig:Concept_MPC} illustrates the concept of the MPC-based iPTM in two distinct scenarios. In the first scenario, as the commercial vehicle approaches the charging station, the prediction horizon encompasses the entire duration from the current time instant to the end time of the charging, which can be segmented into two phases: the driving phase and the charging phase. In the second scenario, as the commercial vehicle arrives at the charging station or waits for the charger, the prediction horizon extends from the current time instant until the end time of the charging, but only consists of one phase, i.e., the charging phase. In both scenarios, the charging time is not known, which is determined as part of the MPC solution. The MPC-based iPTM solves the following discrete-time finite-horizon optimization problem:

\vspace{-4pt}
\begin{equation} \label{eq:MPC_formulation}
\small
\begin{aligned}
& \min_{\substack{\dot{Q}_{bat}\cdot,\\\dot{Q}_{cab}\cdot,\\P_{chg}\cdot,~\Delta t_{2}\cdot}} & & ~~\sum_{i=t}^{t+N_{1}-1} (\frac{\dot{Q}^2_{bat}(i)+\dot{Q}^2_{cab}(i)}{COP(i)}\Delta t_1)~~+\\
&
& &\sum_{i=t+N_{1}}^{t+N_{1}+N_2-1}\{(\frac{\dot{Q}^2_{bat}(i)+\dot{Q}^2_{cab}(i)}{COP(i)}\Delta t_2) +\\
&
& &\alpha (\Delta t_2(i))^2+\beta_1 \epsilon_1^2+\beta_2 \epsilon_2^2 \}, \\ 
& \text{s.t.}
& & SOC(i+1)=SOC(i)+f_{soc}(i)\Delta t_j,~~j\in\{1,2\}\\
&
& & T_{bat}(i+1)=T_{bat}(i)+f_{bat}(i)\Delta t_j,~~j\in\{1,2\}\\
&
& & T_{cab}(i+1)=T_{cab}(i)+f_{cab}(i)\Delta t_j,~~j\in\{1,2\}\\
&
& & SOC_{min}\le SOC(i) \le SOC_{max},\\
&
& & SOC(t+N_1+N_2) = SOC_{targ},\\
&
& & T_{bat,min}\le T_{bat}(i) \le T_{bat,max}+\epsilon_1,\\
&
& & T_{cab,min}\le T_{cab}(i) \le T_{cab,max}+\epsilon_2,\\
&
& & 0 \le \dot{Q}_{bat}(i) \le \dot{Q}_{bat,max},\\
&
& & 0 \le \dot{Q}_{cab}(i) \le \dot{Q}_{cab,max},\\
&
& & \dot{Q}_{bat}+\dot{Q}_{cab}\le\dot{Q}_{max},\\
&
& & 0\le P_{chg}(i) \le P_{chg,max},\\
&
& & 0\le \Delta t_{2}(i) \le \Delta t_{2,max},
\end{aligned}
\end{equation}
The index $j\in\{1,2\}$ is determined as follows:
%
\begin{equation}
    j=\begin{cases}1,~~~\text{\textbf{if}}~~i\leq t+N_1-1,\\
    2,~~~\text{\textbf{if}}~~i \ge t+N_1.\end{cases}
\end{equation}
\vspace{-4pt}

It can be seen from (\ref{eq:MPC_formulation}) that the cost function consists of three terms: i) the energy consumed for battery and cabin cooling, ii) the penalty term of charging time (represented by $\Delta t_2$, as the number of samples in the charging horizon is fixed by $N2$), and iii) slack variables ($\epsilon_1$ and $\epsilon_2$) defining soft constraints of the upper bounds of $T_{bat}$ and $T_{cab}$. Note that the control/decision variables are $\dot{Q}_{bat}$, $\dot{Q}_{cab}$ $P_{chg}$, and $\Delta t_{2}$. 

$f_{SOC}$, $f_{bat}$, and $f_{cab}$ are the dynamic equations (1-3) introduced in Section II. The variables $\Delta t_1$ and $\Delta t_2$ represent the sampling time during the driving and charging phases, respectively, while $N_1$ and $N_2$ represent the number of sampling points during these phases. The prediction horizon length can be calculated as $\Delta t_1 N_1 + \Delta t_2 N_2$, where $\Delta t_1 N_1$ is the remaining time it will take for the vehicle to arrive at the charging station, and $\Delta t_2 N_2$ is the total predicted time spent at the charging station. Note that $N_1$ equals zero once the driving phase is completed and the charging phase begins. 

Note that the MPC formulation (\ref{eq:MPC_formulation}) employs a non-uniform sampling strategy for the prediction horizon. Over the driving phase, the sampling time $\Delta t_1$ is fixed, and the number of samples $N_1$ is calculated based on the remaining time that the vehicle takes before arriving at the charging station. On the other hand, over the driving phase, as the charging time is not predetermined, we fix the sampling number $N_2$, and treat the number of samples $\Delta t_2$ as a control/decision variable, the value of which is determined by solving the optimization problem (\ref{eq:MPC_formulation}). This sampling strategy allows us to impose the terminal condition of $SOC$ at the end of the prediction horizon without knowing the exact time for the vehicle to complete charging.


\section{Simulation Results} \label{section.4}

In this section, we present the numerical results of applying the proposed MPC-based iPTM to a commercial EV, operating at an ambient temperature of $38^oC$. It starts with a low initial $SOC$, drives through an urban route to a fast charging station, and gets charged to a target $SOC$, as shown in Fig.~\ref{fig:Concept_MPC}. 

For the results shown below, we use 0.3 (30\%) and 0.6 (60\%) for the initial and target $SOC$, respectively. The maximum charging power ($P_{chg,max}$) is $80~kW$. To protect the OEM proprietary data, the exact value of $\dot{Q}_{max}$ will not be given in this paper, and the maximum cooling power of the battery and cabin are both $83\%$ of $\dot{Q}_{max}$. The cooling power results shown in this section will be normalized using $\dot{Q}_{max}$. During both the driving and charging phases, the desired range of battery temperature and cabin temperature is $15$ to $35^oC$ and $23$ to $25^oC$, respectively. Note that the range of battery temperature is suggested by~\cite{pesaran2013tools}, and the cabin temperature range is adopted from the temperature setting in the flight of Korean Air~\cite{Korean_air}.

\subsection{Simulation results with accurate preview}

\vspace{-0pt}
\begin{figure*}[h!]
	\begin{center}
		\includegraphics[width=1.2\columnwidth]{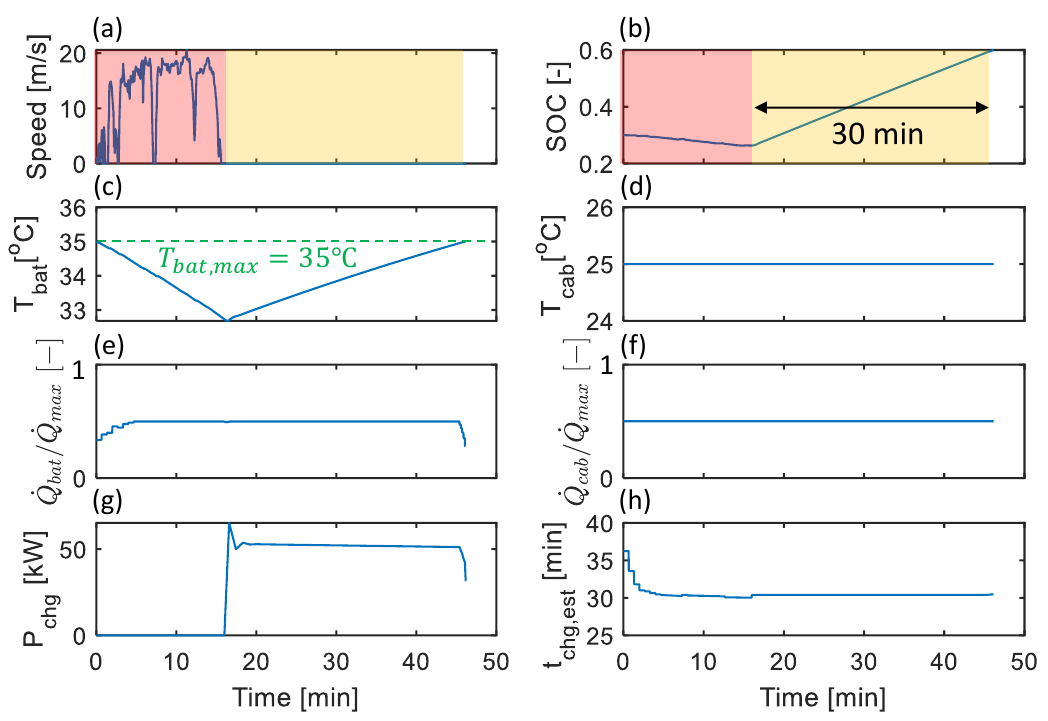}\vspace{-0pt}
		\caption{State and input trajectories with MPC and accurate preview: (a) Vehicle speed, (b) $SOC$, (c) battery temperature, (d) cabin temperature, (e) scaled battery cooling power, (f) scaled cabin cooling power, (g) (negative) battery charging power, and (h) the estimated charging time.}\vspace{-5pt}
	     \label{fig:State_Trajectory_Accurate} 
	\end{center}
\end{figure*}

In order to show the effectiveness of the MPC-based iPTM, we first assume that accurate preview information over the entire prediction horizon is available. In this study, the preview information consists of the vehicle speed profile before arriving at the charging station and the waiting time at the charging station before charging begins. Such an assumption will be relaxed later for a study of the robustness of the algorithm. In this and the following case studies, the budgeted charging time is set to $30~min$.

Fig.~\ref{fig:State_Trajectory_Accurate} presents the state and input trajectories with the MPC. It can be seen that with accurate preview information, the requirement of charging time can be satisfied. After charging for $30~min$, the $SOC$ reaches the required value of 0.6. It can be seen from Fig.~\ref{fig:State_Trajectory_Accurate}-(g) that to meet the desired charging time, a substantial charging power of approximately $50~kW$ is necessary in this case study. However, due to the capacity constraint of the cooling system, battery heat generation during the charging event may drive temperature to the upper constraint. As shown in Fig.~\ref{fig:State_Trajectory_Accurate}-(c), to avoid constraint violations, the controller pre-cools the battery temperature before charging began. Such pre-cooling creates some room for the battery temperature to rise during the charging phase and thus can avoid the throttling of the charge rate.

To conclude, this case study demonstrates the effectiveness of the proposed MPC-based iPTM in enforcing fast charging requirements and power/thermal constraints throughout the driving and charging phases. Nonetheless, such favorable behavior is contingent on  precise preview information, e.g., the vehicle speed, and availability of the charging station. Uncertainties may potentially impede the controller's performance. 

\subsection{Impact of the uncertainties in the preview}

We next consider the impact of the uncertainties in the preview information. As discussed in Sec IV-(B), MPC incorporates the knowledge of the upcoming charging event to pre-cool the battery, thereby allowing large battery charging power to reduce the charging time. The following cases are considered to illustrate the impact of accurately predicting the charging event:

\begin{itemize}
    \item \textbf{Case~I}{:~The charging event is predicted accurately over the prediction horizon,}
    \item \textbf{Case~II}{:~The charging event is not predicted until the vehicle arrives at the charging station.}
\end{itemize}

\begin{figure*}[h!]
	\begin{center}
		\includegraphics[width=1.2\columnwidth]{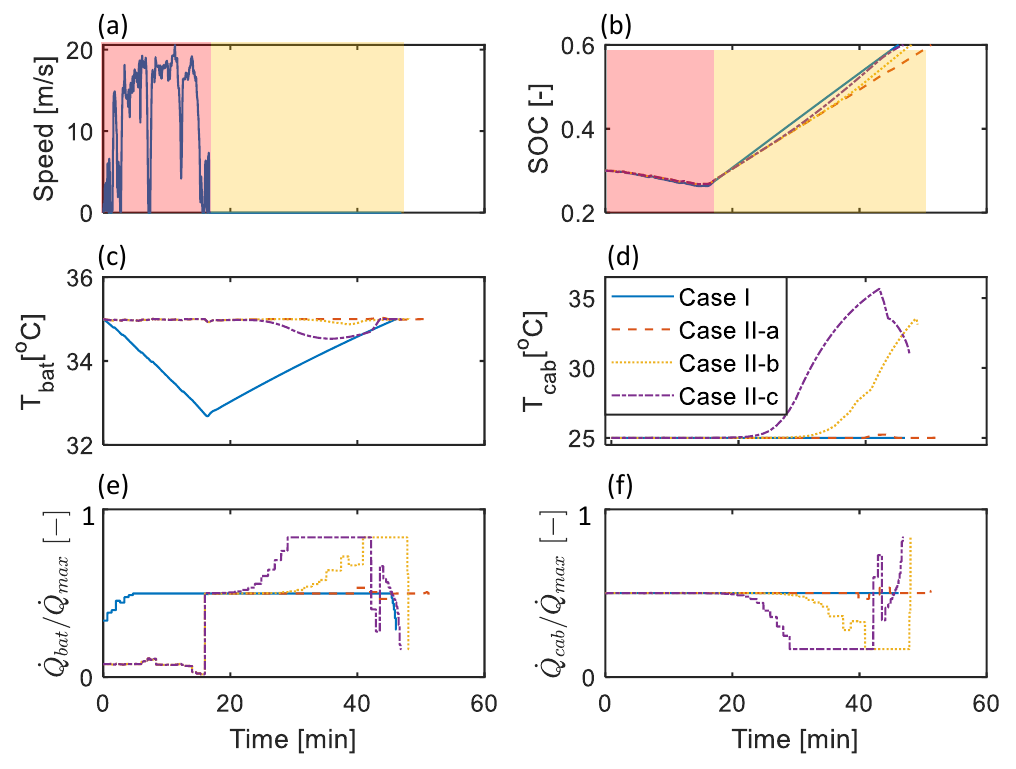}\vspace{-0pt}
		\caption{State and input trajectories of with MPC and preview/without a preview of charging event and different weights: (a) Vehicle speed, (b) $SOC$, (c) battery temperature, (d) cabin temperature, (e) scaled battery cooling power, and (f) scaled cabin cooling power.}\vspace{-10pt}
	     \label{fig:State_Trajectory_Uncertain} 
	\end{center}
\end{figure*}

For Case I, it is assumed that the preview information is known a priori. In Case II, before the vehicle arrives at the station, the charging event is not predicted, and thus, the cost function only includes the first term in (\ref{eq:MPC_formulation}) to minimize the BTM energy consumption. Once the vehicle arrives and starts charging, the same optimization problem in (\ref{eq:MPC_formulation}) will be solved. The comparison of Case I and Case II is displayed in Fig.~\ref{fig:State_Trajectory_Uncertain}. Note that for Case II, a different set of penalty weights is defined in Table~\ref{tab:beta} from Case II-a to II-c.  

\vspace{-0pt}
\vspace{-00pt}

Unlike Case I, which benefits from accurate preview information, in Case II MPC does not account for the upcoming charging event over the prediction horizon and therefore does not conduct pre-cooling on the battery. This uncertainty leads to performance degradation of the controller, and different settings of penalty weights ($\beta_1$ and $\beta_2$) influence the priority of the multi-objective MPC. 

\vspace{-2pt}
\begin{table}[h!]
\centering 
{\caption{Penalty weights of $\beta_1$ and $\beta_2$. \label{tab:beta}}}\vspace{-5pt}

\begin{tabular}{ccccc}
 & Case I & Case II-a  & Case II-b & Case II-c\\
\hline
$\beta_1$  & 1e11 & 1e11 & 1e11 & 1e11\\
$\beta_2$  & 1e10 & 1e10 & 1e5 & 1e3\\
$t_{chg}$  & 30.0 & 35.6 & 32.5 & 31.2\\
$[min]$ & & & &\\
$CV$  & 0 & 44 & 3680 & 8740\\
$[^oC \cdot sec]$ & & & &\\
\hline
\end{tabular}\vspace{-2pt}
\end{table}

Note that $\beta_1$ is always set to a large value to ensure that the battery temperature constraint is enforced, while different values of $\beta_2$ are selected, representing varying levels of relaxation for the cabin temperature constraint. Moreover, the charging time ($t_{chg}$) and accumulated cabin temperature constraint violation ($CV$) are summarized in Table~\ref{tab:beta}.

\begin{figure}[h!]
	\begin{center}
		\includegraphics[width=0.8\columnwidth]{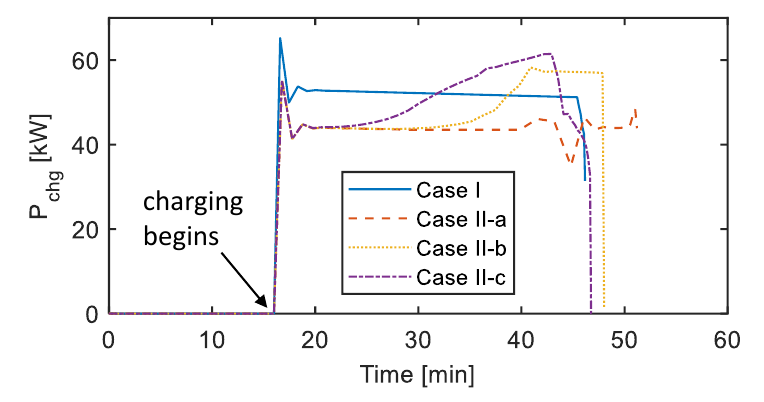}\vspace{-10pt}
		\caption{The time histories of the (negative) charging power in different cases.}\vspace{-8pt}
	     \label{fig:P_chg_uncertain} 
	\end{center}
\end{figure}

It can be seen that for Case II, the MPC performance is sensitive to the settings of $\beta_2$, and there is a trade-off between the charging time and cabin temperature constraint violation. As the value of $\beta_2$ is reduced, the accumulated constraint violation of cabin temperature increases, while the charging time reduces. This is because of the constraint on cooling capacity. For Case II-a, with large values of $\beta_1$ and $\beta_2$, MPC tends to enforce the constraints for both battery and cabin temperature. However, because no pre-cooling is conducted in advance, there is no room for the battery temperature to rise. Therefore, compared with Case I, in Case II-a the charging power needs to be reduced to avoid constraint violation, which causes an increase in total charging time. For Case II-c, as $\beta_2$ is reduced, the strictness of the cabin temperature constraint is relaxed, which allows cabin temperature to rise above its upper bound. Therefore, the cabin cooling power can be reduced and the battery cooling power can be increased, allowing for a larger charging as shown in Fig.~\ref{fig:P_chg_uncertain}. 

This case study illustrates that failure to predict the charging event in advance results in performance degradation and the MPC-based iPTM has to compromise either on the charging time or the enforcement of constraints.

\subsection{$SOC$-aware penalty weight adjustment strategy}

The case study in Sec.IV-C shows the impact of the uncertainty in predicting the charging event with constant penalty weights. In this section, we explore the time-varying penalty weights to enhance the performance of the MPC-based iPTM strategy. An adjustment strategy is developed to update the penalty weights in real-time based on the current state of $SOC$. 

Two assumptions are made in this study. First of all, it is assumed that during the charging phase, the drivers and the passengers leave the vehicle. This assumption implies that any constraint violation of cabin temperature during the charging phase will not affect the comfort of the drivers and passengers. Secondly, we assume that the drivers and passengers return to the vehicle immediately after the charging is completed. Therefore, the final cabin temperature will be used as a metric to evaluate the comfort level upon their return. 

With these two assumptions, a time-varying weighting strategy is proposed to enhance the charging performance in situations where the charging event is not correctly predicted. This strategy exploits the insight that the cabin temperature constraint can be relaxed during the early phases of charging when there is no occupants in the vehicle but should be tightened as the charging nears completion. This idea is implemented with the penalty weight ($\beta_2$) given by:

\begin{gather}\label{eq:SOC_awareness}
\beta_2=F(SOC)=\beta_{0}10^{b\frac{SOC-SOC_{min}}{SOC_{targ}-SOC_min}}
\end{gather}

It can be seen that $\beta_2$ takes value between $\beta_{0}$ (when it starts charging) and $10^b\beta_{0}$ (when $SOC$ approaches the target). Therefore, by using this equation, the tightness of the cabin temperature constraint is adjusted. Hence a new case study is defined as:

\begin{itemize}
    \item \textbf{Case~III}{:~The charging event is not predicted until the vehicle arrives at the charging station. $\beta_2$ is time-varying, following eq.~(\ref{eq:SOC_awareness}) with $\beta_{0}=10$ and $b=10$.}
\end{itemize}

\begin{figure}[h!]
	\begin{center}
		\includegraphics[width=0.7\columnwidth]{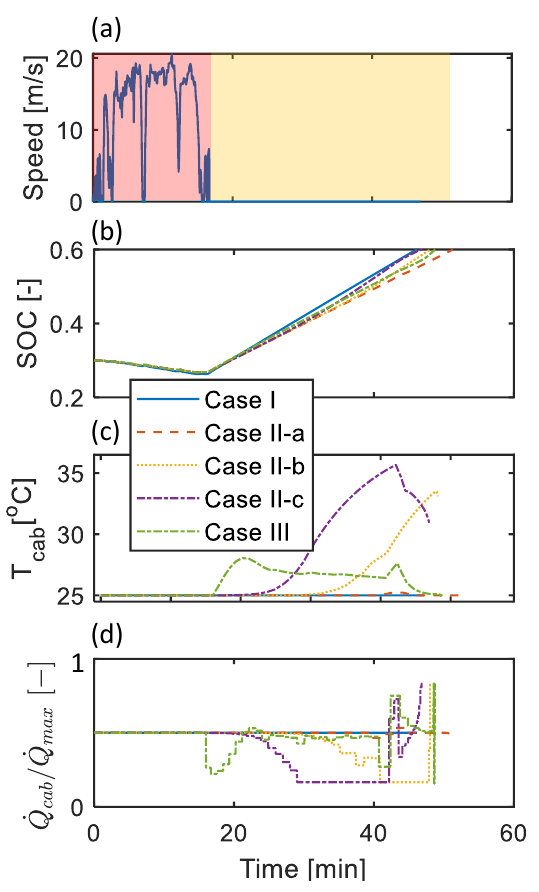}\vspace{-10pt}
		\caption{State and input trajectories with MPC when assuming the preview information is accurate: (a) Vehicle speed, (b) $SOC$, (c) cabin temperature, and (d) scaled cabin cooling power.}\vspace{-8pt}
	     \label{fig:State_Trajectory_SOC} 
	\end{center}
\end{figure}

The comparison of trajectories with MPC in Case I-III is presented in Fig.~\ref{fig:State_Trajectory_SOC}. It can be seen from Fig.~\ref{fig:State_Trajectory_SOC}-(c) that the cabin temperature rises above the upper bound ($25^oC$) at the beginning of the charging and cools down when charging nears completion. The charging time, and final cabin temperature in all cases are summarized in Table.~\ref{tab:results2}.

\vspace{-2pt}
\begin{table}[h!]
\centering 
{\caption{Settings of penalty weights of $\beta_1$ and $\beta_2$. \label{tab:results2}}}\vspace{-5pt}

\begin{tabular}{ccccc}
 & Case I & Case II-a  & Case II-c & Case III\\
\hline
$t_{chg}$  & 30.0 & 35.6 & 31.2 & 32.9\\
$[min]$ & & & &\\
$T_{cab,final}$  & 25.0 & 25.0 & 31.0 & 25.0\\
$~[^oC]$ & & & &\\
\hline
\end{tabular}\vspace{-2pt}
\end{table}

It can be seen that while both Case II-a and Case III successfully enforce the final cabin temperature constraint, Case III outperforms Case II-a by reducing charging time by $2.7~min$. On the other hand, although Case II-c requires less charging time than Case III, the final cabin temperature is higher, when occupants return to the vehicle after charging is completed. Consequently, the proposed time-varying weighting strategy improves the charging performance and enforces the thermal constraints in the presence of uncertainty. Such benefits are achieved by adjusting the strictness of the constraint during the charging phase.

\section{Summary and Conclusions} \label{section.5}

In this paper, a multi-objective model predictive control (MPC) framework is developed to optimize the fast charging performance through integrated power and thermal management (iPTM). The controller takes into account the coupling of the battery and cabin thermal management. Due to the uncertain charging time, the proposed MPC-based iPTM utilizes a non-uniform sampling over a prediction horizon and a shrinking horizon to achieve the desired performance. The simulation results demonstrate the effectiveness of the proposed MPC-based iPTM in achieving fast charging while enforcing power/thermal constraints throughout the driving and charging phases when accurate preview information is available. However, the uncertainty analysis shows that the failure to predict the charging event leads to performance degradation, and the controller needs to compromise either on the charging time or the cabin temperature at the end of the charging event. Moreover, to enhance the performance in the presence of uncertainty, a time-varying weighting strategy was proposed to leverage the different thermal priorities and adjust the penalty weights in the cost function. This strategy updates penalty weights based on the current value of the battery $SOC$. The simulation results show that at the same value of the final cabin temperature, the proposed strategy reduces the required charging time by adjusting the cabin priority during the charging phase.

\section*{ACKNOWLEDGMENT}
Julia Buckland Seeds, Hao Wang, and Connie Qiu from Ford Motor Company are gratefully acknowledged for their technical comments during the course of this study.


\bibliographystyle{IEEEtran}
\bibliography{Reference.bib}

\end{document}